\documentclass[aps,twocolumn,prb,superscriptaddress,floatfix,showpacs]{revtex4}
\usepackage{amsmath}
\usepackage{graphicx}
\usepackage{subfigure}
\usepackage{color}
\usepackage{epstopdf}

\newcommand{\lam}{\lambda}
\newcommand{\del}{\delta}

\newcommand{\ba}{\begin{eqnarray}}
\newcommand{\ea}{\end{eqnarray}}
\newcommand{\ct}{\cite}

\newcommand{\la}{\lambda}
\newcommand{\bi}{\bibitem}
\newcommand{\be}{\begin{equation}}
\newcommand{\ee}{\end{equation}}
\newcommand{\al}{\alpha}

\newcommand{\non}{\nonumber}
\newcommand{\de}{\delta}

\begin{document}
\title{One and two-dimensional quantum models: quenches and the scaling of irreversible entropy}
\author{Shraddha Sharma}

\author{Amit Dutta}
\affiliation{Department of Physics, Indian Institute of Technology, 208016, Kanpur}

\begin{abstract}

Using the scaling relation of the ground state quantum fidelity, we propose the most generic  scaling relations  of  the irreversible work (the  residual energy) of a closed  quantum system  at absolute zero temperature when one of the parameters of its Hamiltonian is suddenly changed; we  consider two extreme limits namely, the heat susceptibility limit and the thermodynamic limit. It is then argued that  the irreversible entropy generated for a thermal quench at  low enough temperatures when the system
is initially in a Gibbs state, is likely to show  a similar  scaling behavior.  To illustrate this proposition, we consider zero-temperature and thermal quenches in one and two-dimensional Dirac Hamiltonians  where the exact estimation of the irreversible work   and the irreversible entropy is indeed
possible. Exploiting these exact results, we  then establish: (i) the irreversible work at zero temperature indeed shows an appropriate scaling in the thermodynamic limit;  (ii) the scaling of the irreversible work in the 1D Dirac model at zero-temperature shows  logarithmic corrections to the scaling which is
a signature of a marginal situation. (iii) Furthermore, remarkably the logarithmic corrections do indeed appear in the scaling of the entropy generated if temperature is low enough while disappears for high temperatures. For the 2D model, no such
logarithmic correction is found to appear.

\end{abstract}

\maketitle


\section{Introduction}

Studying emergent thermodynamics following  a
quench of a closed quantum system initially in equilibrium with a heat reservoir  has attracted the attention of the scientists in recent years \cite{talkner07,talkner07_work,deffner10,dorner12}. These studies are also
important from the viewpoint  of non-equilibrium fluctuation theorems \ct{crooks99,tasaki00,talkner07,talkner07_work} and quantum 
Jarzynski equality \ct{jarzynski97}. Treating a sudden quench (i.e., a change) of a parameter of the Hamiltonian
as a  thermodynamic transformation and decoupling the system from the bath so that the evolution of the system following
a quench is perfectly unitary, a generalized second law (or Clausius inequality) for an  isolated quantum system has been posited {with the entropy generated in the process being defined  in terms} 
of the irreversible work done in the process of quenching \ct{deffner10, dorner12,sotiriadis13}. These studies have been
generalized to the context of Bures metric \ct{zanardi07_bures} and Uhlmann fidelity \ct{deffner131}, single
qubit interferrometry \ct{dorner13}, open quantum systems \ct{deffner132} and inner friction of quantum thermodynamic
processes \ct{plastina14}.

In the above studies, the quantum system is initially prepared in a mixed state in equilibrium with a heat bath which
is then decoupled and the system is subjected to a sudden change (or a slow ramp dictated by  a protocol described by a unitary operator $\hat U$)  of one of the  parameters. Due to the stochastic nature of
the work done for a finite system (which is \textit{not} an observable \ct{talkner07_work}), one makes resort to
the characteristic cumulant-generating function. To elaborate,  let us consider a system described by the Hamiltonian  $H(\lam)$ which
is at the quantum critical point (QCP) at $\lam=0$ \ct{Sachdev,suzuki13,dutta10}; we assume that  at an initial time $t=0$ the system  is kept in contact with a heat bath at an
inverse temperature $\beta$ with an  initial value of $\lambda=\lambda_0$.  The system is then decoupled from the bath and the
parameter $\lam$ is tuned  from the initial value $\lam_0$  to the  final value $\lam_f =\la_0 +\del$, so that the subsequent  temporal evolution of
the system is entirely dictated by the unitary operator $\hat U$. Implementing the analytical continuation $s \to -it$, one defines
the  characteristic function 
\ba
G(s)={\rm Tr} [{\hat U}^{\dagger} e^{sH(\lambda_f)} \hat {U} e^{-sH(\lambda_0)}\rho_0(\lam_0)];
\label{Gs1}
\ea
\noindent here, $\rho_0(\lam_0)=e^{-\beta
H(\lambda_0)}/Z(\lambda_0)$ is the density matrix characterizing the initial Gibbs state and 
$Z(\lambda_0)$ is the corresponding partition function.  For an instantaneous  quench from $\la_0$
to $\la_f$,  $G(s)$ takes a simpler form:
\ba
G(s)={\rm Tr} [e^{sH(\lambda_f)} e^{-sH(\lambda_0)}\rho_0 (\lam_0)].
\label{Gs1_sudden}
\ea
At this juncture, it should be noted   that the characteristic function $G(s)$ closely resembles the mixed state Loschmidt
echo \ct{zanardi07_echo} which has been extensively studied in the context of zero-temperature non-equibrium
dynamics of quantum critical systems \ct{quan06,rossini07,sharma12,mukherjee12,nag12} and 
associated dynamical phase transition \ct{heyl13}. Moreover, $G(s)$  is related (through
the inverse Fourier or Laplace transform) to the distribution function $P(W)$ of work ($W$), which is usually
characterized by an edge singularity at zero-temperature \ct{silva08,gambassi11, smacchia13}.

Focussing on the sudden quenching case,  the average work performed  is defined through the first cumulant of the characteristic function, as shown:

\begin{equation}
 \langle W\rangle=dG(s)/ds|_{s=0}={\rm Tr}\left(H(\lam_f)\rho_0\right)-{\rm Tr}\left(H(\lam_0)\rho_0\right).
\label{Wav}
\end{equation}

\noindent To arrive at the generalized second law, we note that for a non-equilibrium process, the average work  $\langle W\rangle$ always exceeds the free energy difference ($\Delta F$) between the initial and final equilibrium states (both with inverse
temperature $\beta$) with parameters $\la_0$
and $\la_f$, respectively. One can define the irreversible work through the
difference 
$W_{\rm irr}^T =\langle W\rangle-\Delta F$. The heat exchange with the bath for a closed quantum system during the
evolution being zero, the irreversible entropy generated in the process satisfies the relation $\Delta S_{\rm irr}=\beta W_{\rm irr}^T$.
In the subsequent discussion, we shall drop the superscript $T$ and use $W_{\rm irr}$ to denote the finite temperature 
irreversible work.

For a sudden quench of magnitude $\del$ at zero-temperature, when $\Delta F \rightarrow (E_g(\lam_f) - E_g (\lam_0))$,   Eq.~(\ref{Wav}) reduces to

\be  
W_{\rm irr}^{T=0} = \langle\psi_g(\lam_0)|H(\lam_f)|\psi_g (\lam_0)\rangle-E_g(\lam_f);
\label{eq_wirr_zero}
\ee
here, $|\psi_g(\lam_0)\rangle$ is the ground state wave function corresponding to the initial  Hamiltonian 
$H(\lam_0)$ and $E_g(\lam_i)$$ (E_g (\lam_f))$ is the ground state energy of $H(\la_i)$ $(H(\lam_f))$. Clearly,  $W_{\rm irr}^{T=0}$ is the same as  the residual energy  (excess energy above the ground state) and hence should follow an identical scaling relation 
for a sudden quench obtained using an adiabatic perturbation theory as reported in the literature \ct{degrandi10, polkovnikov11,dutta10,dziarmaga11}.
It is noteworthy that very recently, the quantity $W_{\rm irr}^{T=0}$ has been studied  establishing
a connection to the  heat  susceptibility (for $\de\to 0$) in the context of quantum latency  which detects the order of a quantum phase
transition (QPT) \ct{mascarenhas14}.
 
Let us now ask the question how does the proximity to a QCP influence the behavior of $\Delta S_{\rm irr}$ and explore its scaling
behavior in terms of the deviation from the QCP ($\la$), the magnitude of the quench $\de = \la_f - \la_0$ and 
the system size $L$.  In Ref. [\onlinecite{dorner12}], it was shown how the form of $G(s)$ given in Eq.~(\ref{Gs1}) can be
exploited to establish the Tasaki-Crooks relation and the Jarzynski equality for a closed quantum system.  Taking the
example of a 1D quantum Ising model, it has also  been shown that when $\la$
is varied keeping $\de$ small and $L$ finite, $\Delta S_{\rm irr}$ shows a sharp peak at the QCP if $\beta$ is large,   while
on the contrary in the high temperature situation ($\beta \to 0$),  $\Delta S_{\rm irr}$ gets broadened.  One therefore concludes that the peak in $\Delta S_{\rm irr}$ is an indicator of a QPT  even at a low but finite temperature.  

In this paper, we shall {first consider $W_{\rm irr}^{T=0}$ and  inspired by the scaling of the quantum fidelity 
explore its scaling behavior for a sudden quench of magnitude $\de$ in the vicinity of a QCP characterized by the associated quantum critical exponents  in different limits $\la$, $\de$ and $L$;}
both the situations when $\de$ can be treated as a perturbation \ct{mascarenhas14} or it is a scaling variable will be probed.  
We shall then address the question whether $\Delta S_{\rm irr}$ generated for such a quench exhibit similar scaling relations.
To the best of our knowledge, our work is the first attempt that generalizes the derivation of $\Delta S_{\rm irr}$ to a 
higher dimensional system where an exact analytical calculation is possible. The exact analytical form enables us to derive
the scaling of $W_{\rm irr}^{T=0}$ and $\Delta S_{\rm irr}$  when $\del \to 0$ and also for finite $\delta$.

%

The paper is organized in the following manner: the scaling relation of $W_{\rm irr}^{T=0}$ and $ \Delta S_{\rm irr}$ are proposed
 in Sec. \ref{sec_scaling}. 
In Sec. \ref{sec_dirac_marg}, we shall invoke upon
the one dimensional (1D) and two dimensional (2D) Dirac Hamiltonians to illustrate the scaling relations proposed earlier.
The advantage of using a Dirac Hamiltonian is the inherent $2 \times 2$ nature of the same rendering it integrable in
all dimensions; in addition, 1D Dirac
Hamiltonian, as we shall show below, provides an ideal example of a marginal situation where we find the signature of
logarithmic corrections to the scaling of $W_{\rm irr}^{T=0}$  which persist in $\Delta S_{\rm irr}$ at  low but finite temperatures and disappears in the 2D case.  In the 2D case,
in contrast to the 1D case, $\Delta S_{\rm irr}$ does not show a sharp peak at the QCP and hence can not be treated
as an ideal indicator of a QPT. Concluding comments are presented in Sec.~\ref{sec_conc}.

\section{scaling relations of $W_{\rm irr}$ and $ \Delta S_{\rm irr}$}
\label{sec_scaling}

Before embarking upon the study of $\Delta S_{\rm irr}$ at finite temperature, we piece together
the scaling of the defect density and the residual energy at zero temperature  following a sudden quench considering different
limits: we first consider   the situation when $L$ is the largest length scale of the problem which implies $L  \gg \de^{-\nu},
\la^{-\nu}$, 
 where $\nu$ is the correlation length exponent associated with the QCP; clearly,  in this case the parameter $\delta$ is finite and hence  can not be
 treated as a perturbation. 
 We shall arrive at the scaling relation of $W_{\rm irr}^{T=0}$  using heuristic arguments inspired
 by the scaling of the ground state  quantum
fidelity \ct{zanardi06,venuti07,gu10},  $F(\la,\de) = |\langle \psi_g(\la_0)| \psi_g(\la_0 + \de)\rangle|$ given by \ct{rams11a}
\be
  \ln F = -L^d |\de|^{\nu d} {\cal F}\left(\frac {\la}{\de}\right).
  \label{logF}
 \ee
where ${\cal F}$ is the corresponding scaling function, $\nu$ is the correlation length exponent associated with the QCP
and $d$ is the spatial dimension; $L$ being the largest length scale of the problem, this limit is known
as the \textit{thermodynamic limit}. 
Let us first consider the situation,  $\la \ll  \de$, which means $\del^{-\nu} \ll \la^{-\nu}$, i.e., the correlation length $\xi \sim \lam^{-\nu}$ is larger
than the length scale associated with the parameter $\del$, so effectively the system  is close to the QCP.
Using the finite size scaling argument \ct{fisher72},
we expect  $\ln F$ to  scale  with $\de$.  
 Demanding that right hand  side of Eq.~\eqref{logF} is dimensionless, we
find a  characteristic momentum scale $\hat k \sim \de^{\nu}$ (since we know that $\de^{-\nu}$ has the dimension of length). Using the phase space argument, one then finds that the defect 
(or quasi-particle) density ($n$)  scales as ${\hat k}^d \sim \de^{\nu d}$. (For $\nu=d=1$, it has been proved \cite{rams11b} that one has $n\sim |\de|$.)  We note that as predicted in Ref. [\onlinecite{degrandi10}], the
scaling of defect density is different from the residual energy {for a sudden quench in the vicinity of} the QCP; in this case,  each excitation carries energy $\de^{\nu z}$, where $z$ is the dynamical exponent,  resulting in the scaling  $W_{\rm irr}^{T=0} \sim \de^{\nu(d+z)}$ \ct{degrandi10,gritsev09}.  

We now consider  the other limit,  when $\la \gg \de$, {(implying $ \la^{-\nu} \ll \de^{-\nu}$, i.e., away from the QCP),  $\la$ is
expected to play the role of the scaling variable}. The quantum
fidelity defined in  Eq.~(\ref{logF}) is expected to scale as  $ \ln F = -L^d \de^2 \la^{\nu d -2} $; in a similar spirit  identifying  the characteristic momentum 
scale,  it is straightforward to arrive at the scaling of the  defect density  as $n \sim {\hat k}^d \sim  \de^2 \la^{\nu d -2}$ which has already been established  for the quantum Ising case \ct{rams11b} ($\nu=d=z=1$) enabling us  
to arrive at the scaling relation,  $W_{\rm irr}^{T=0} \sim \de^2 \la^{\nu d  + \nu z -2}= \de^2 \la^{-\alpha}$ where $\al$ is the 
corresponding ``specific heat" exponent. 

We have illustrated  above how  the scaling of  fidelity  can lead to the scaling of $W_{\rm irr}^{T=0}$ in the thermodynamic
limit. Concerning the scaling of fidelity,  it is well-known that one can switch from  the thermodynamic limit   to the fidelity susceptibility ($\chi_F(\la)$) limit \ct{schwandt09,degrandi10,gritsev09,mukherjee11} (where the notion of the  fidelity susceptibility is meaningful) 
when $\de^{-\nu}\gg L$.  Continuing {along} the same line of arguments, one can define the heat susceptibility $\chi_E$ \ct{degrandi10} such that  $W_{\rm irr}^{T=0} \sim \de^2 \chi_E$, when $\de^{-\nu}$ (i.e, $\de \to 0$) is the largest length scale. In this limit, one expects a scaling $W_{\rm irr}^{T=0} \sim \de^2 \la^{-\alpha}$ away from the QCP 
and on the other hand, close to the QCP,  $L$ plays the role of the scaling variable in lieu of $\la$. To summarize, we have:

\medskip

\noindent {\bf{Heat Susceptibility limit:}} $\de^{-\nu}$ is the largest length scale and $\de (\to 0)$ can be treated as a perturbation\\
\ba
 W_{\rm irr}^{T=0}/L &\sim& \del^2 \lam^{\nu(d+z)-2}~~~ \del^{-\nu}>L>\lam^{-\nu} \nonumber\\
&\sim& \del^2 L^{2/{\nu}-(d+z)}~~~ \del^{-\nu}>\lam^{-\nu}>L
\label{scal1_susc}.
\ea
The crossover between the limits close and away from the QCP occurs when $L \sim  \lam^{-\nu}$.

\noindent{\bf{Thermodynamic limit}}: $L$ is the largest length scale:\\
\ba
 W_{\rm irr}^{T=0}/L &\sim& \del^2 \lam^{\nu(d+z)-2}~~~~ L>\del^{-\nu}>\lam^{-\nu} \nonumber\\
&\sim&  \del^{\nu(d+z)}~~~~~~~~~  L>\lam^{-\nu}>\del^{-\nu}
\label{scal_thermo}.
\ea
Similarly here the  crossover from one scaling to the other occurs when $\lam \sim \del$. As explained above the crossover
from the thermodynamic limit to the susceptibility limit occurs when $\del^{-\nu} \sim L$. {At this point, it would
be instructive to note that what we present above happens to be  the most general scaling relations of $W_{\rm irr}^{T=0}$. If
the system is precisely at the QCP ($\lam^{-\nu} \to \infty$), one arrives at the scaling relations (with $L$ or $\de$) proposed in Ref.~[\onlinecite{degrandi10}] by setting $\la =\de$ in the scaling relations and switching  $\la$ and $\de$ in the
corresponding conditions in {Eqs.~(\ref{scal1_susc}).}

We would like to emphasize here that when the combination $\nu(d+z)$ exceeds $2$, the scaling
of $W_{\rm irr}$ is non-universal. In the marginal case when $\nu(d+z)=2$,  one encounters additional   logarithmic corrections to the scaling and one needs to introduce an upper cut-off in the momentum scale $k_{\rm max}$
as we shall illustrate below for the 1D Dirac model.

We shall now proceed to analyze the scaling of $\Delta S_{\rm irr}$ generated  following a sudden change of $\lam$ by an amount $\delta$
 while the system was initially in thermal equilibrium with a heat bath at a temperature $T$.
Question we raise  is that wether $\Delta S_{\rm irr} =\beta W_{\rm irr}$, is likely to show a similar
scaling relation for such a thermal quench.  {Using a perturbative expansion of the free energy (corresponding
to the final parameter) valid in the limit $\del \to 0$, a scaling $\Delta S_{\rm irr} \sim \de^2 \la^{-\al}$, has been proposed \ct{sotiriadis13}; obviously, a similar
expansion is not possible in other limits mentioned in Eqs. (\ref{scal1_susc}) and (\ref{scal_thermo}).   In the finite temperature case, the system is initially in a mixed state.
Nevertheless, for low enough  temperatures the ground state of the Hamiltonian is maximally populated {and} one would therefore expect $\Delta S_{\rm irr}$ to satisfy a similar scaling as  $W_{\rm irr}^{T=0}$;
on the contrary, these scaling relations
should disappear at higher temperatures when the initial  state deviates significantly from the ground state.
 Our aim here is to show that this  indeed is the case in all the limits not only for $\del \to 0$.} Furthermore,  we shall illustrate how to arrive at these
scaling relations in both the susceptibility limit and the thermodynamic limit for an integrable Hamiltonian. Finally,  the question whether  at a low temperature  $\Delta S_{\rm irr}$ always shows a sharp peak at the QCP will be addressed.

\section{Scaling in Dirac Hamiltonians and role of marginality}

\label{sec_dirac_marg}

In this section, we shall derive the scaling relations of $W_{\rm irr}^{T=0}$ and address the issues raised in the previous section concerning the scaling of $\Delta S_{\rm irr}$ using
the example of 1D and 2D Dirac Hamiltonians which are  very important from theoretical point of view in quantum condensed matter systems. 
For example, we take a 2D massive Dirac Hamiltonian
$\hat{H}_D = \int \! d{\bf x} \; \widehat{\Psi}^{\dagger}({\bf x}) \Big[ m \hat{\sigma}_z -i\hbar v_F (\hat{\sigma}_x \partial_x + \hat{\sigma}_y \partial_y ) \Big] \widehat{\Psi}({\bf x})$,
where $\widehat{\Psi}({\bf x})$ is a two-component spinor field operator describing,  the effective low-energy degrees of
freedom for electrons on a honeycomb lattice with unequal sublattice potentials around a single Dirac point (i.e., single valley graphene Hamiltonian) with Fermi velocity $v_F$ (set equal to unity below)\cite{castroneto09}. Variations of this model are ubiquitous in the field of topological insulators \cite{Kane_RMP10,Zhang_RMP11}. The edge states
in a 2D topological insulator are described by an effective 1D Dirac Hamiltonian, whereas its bulk states are described by a 2D Dirac Hamiltonian and the  QCP separating the gapped to gapless phases is a 2D massless Dirac point. 
These models have turned out to be immensely useful  in the studies of the Kibble-Zurek scaling \ct{dutta10_epl}, sudden quenches \ct{mukherjee_11,patel12}, fidelity susceptibility and thermodynamic
fidelity \ct{mukherjee_thermo_12}, Loschmidt echo \cite{patel131} and periodic steady state reached through a sinusoidal variation of the mass term \ct{sharma14}.
Considering the massive 2D Dirac Hamiltonian 
one can rescale the units appropriately to obtain the $2 \times 2$ Hamiltonian describing the system
close to a single valley as,

\begin{equation}
 H_{\small{2D}} (m)= 
 \left(
 \begin{array}{cc}
  m & k_x-i k_y  \\
 k_x+i k_y & -m   \\
 \end{array}
 \right),
 \label{2DH}
 \end{equation} 
here,  $m$ is the Dirac mass and the vector $\vec{k}$ is measured with respect to the corners of the Brillouin zone. The scaling relations obtained for the  1 D massive Dirac  Hamiltonian:

\begin{equation}
 H_{\small{1D}} (m)= 
 \left(
 \begin{array}{cc}
  m & k  \\
 k & -m   \\
 \end{array}
 \right),
 \label{1DH}
 \end{equation} 
 are to be compared with the corresponding 2D case. Both the 1D and 2D models exhibit a  QCP where the mass term $m$ vanishes and  the dispersion at the QCP is linear. These QCPs are then characterized by the critical  exponents $\nu=z=1$. In the sense of universality $H_{\small{1D}}$ is identical to a 1D quantum Ising chain; the later can be  recast to the  form given in Eq.~(\ref{1DH}) with $m$ being
 the deviation of the transverse field from its quantum critical value.
The 1D case is conspicuous in the sense that we have $\nu(d+z)=2$. As a result, one expects a logarithmic correction to the  scaling of 
$W_{\rm irr}^{T=0}$ and remarkably this logarithmic scaling persists in $\Delta S_{\rm irr}$ even at low temperature. On the contrary, no
such logarithmic singularity is observed in the 2D case.

Using Hamiltonians (\ref{2DH}) and (\ref{1DH}), we shall consider a sudden quench of the mass term $m$ (which
will still be denoted by the parameter $\lam$ for consistency) from an  initial value $\lam$ to a final value $\lam+\del$
and  calculate the exact analytical expressions for $W_{\rm irr}^{T=0}$ and $\Delta S_{\rm irr}$; this would  in turn lead to
the scaling relations of these quantities both in the susceptibility as well as the thermodynamic limit.

Using the eigenstates of  1-D Dirac Hamiltonian given by

\begin{equation}
 |{\psi_k}^{\pm} (\lam_0)\rangle_{\small{1D}}= \frac{k}{\sqrt{(\lam_0-\sqrt{k^2+\lam_0^2})^2+k^2}}
 \left(
 \begin{array}{cc}
  \frac{\lam_0-\sqrt{k^2+\lam_0^2}}{k} \\
  1 \\
 \end{array}
 \right),
 \label{S1D}
 \end{equation} 
with corresponding eigen-energies $E(\lam_0)_{1D}=\pm\sqrt{k^2+\lam_0^2}$ in Eq.~(\ref{eq_wirr_zero}),  the expression for $W_{\rm irr}^{T=0}$ can be derived in a closed form

\begin{widetext}
\ba
{W_{\rm irr}}^{T=0}_{1D} &=& \frac{1}{2} [k_{\rm max}\sqrt{k_{\rm max}^2+(\lam+\del)^2} -k_{\rm max}\sqrt{k_{\rm max}^2+\lam^2}  +(\frac {2\pi}{L})\sqrt{(\frac {2\pi} {L})^2+\lam^2} \nonumber\\
&-&(\frac {2\pi} {L})\sqrt{(\frac {2\pi}{L})^2+(\lam+\del)^2} 
+ (\lam +\del)^2  \log \left(\frac{k_{\rm max}+\sqrt{k_{\rm max}^2+(\lam+\del)^2}}{\frac {2\pi} {L}+\sqrt{(\frac {2\pi} {L})^2+(\lam+\del)^2}}\right)\nonumber\\
&+& \lam(\lam+2 \del) \log\left(\frac{\frac {2\pi} {L}+\sqrt{(\frac {2\pi} {L})^2+\lam^2}}{k_{\rm max} +\sqrt{k_{\rm max}^2+(\lam+\del)^2}}\right)].
\label{Wirr_anl}\ea
 \end{widetext}
Here, we have converted the sum over $k$ to the integral using a periodic boundary condition with the step size $2\pi/L$; $k_{\rm max}$  plays the role of the upper cut off  in the momentum scale, i.e.,  $k_{\rm max}^{-1}$ which is the shortest length scale of the problem (i.e., the lattice spacing). This arises
because of the fact that  the integral in Eq.~(\ref{Wirr_anl}) can not be extended to $k \to \infty$ and to avoid divergence
one needs to introduce a cut-off.  Furthermore, as is evident in the expression in Eq.~(\ref{Wirr_anl}),  the presence of $k_{\rm max}$ in fact renders
the argument of logarithm dimensionless.  Similarly, one
can derive an exact analytical form of the $W_{\rm irr}^{T=0}$ in the 2D case also. {In Fig.~(\ref{Fig:wirr_1d_2d}), we show
the variation of $W_{\rm irr}^{T=0}$ as a function of $\la$ systematically in different limits. We note that when plotted
on  the same scale, there is a significant peak at the QCP in the thermodynamic limit while the peak almost disappears
in the susceptibility limit.}

\begin{figure}[h!]
\includegraphics[width=8cm]{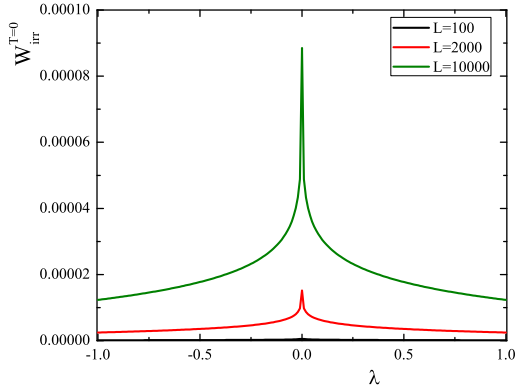}
\caption{The plot of  $W_{\rm irr}^{T=0}$ in different limits showing peak at the QCP ($\lam=0$) for a 1D Dirac Hamiltonian There is 
no visible peak in the susceptibility limit  ($L=100$, bottom most)  and there is a sharp peak in the thermodynamic limit ($L=10000$,
topmost). In the intermediate case, $L=2000$, there is a significant peak. Here,
$\del =0.001$ and $k_{\rm max}=2\pi$ in all  the cases.}
\label{Fig:wirr_1d_2d}
\end{figure}

\begin{figure}[h!]
\includegraphics[width=7.5cm]{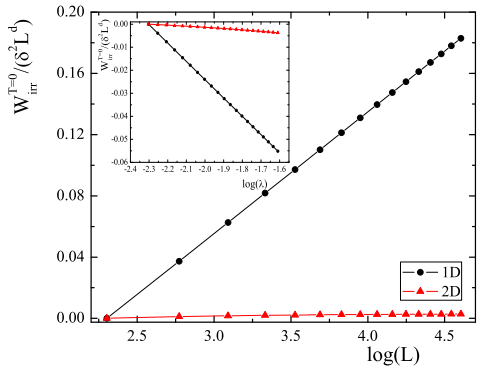}
\caption{The logarithmic dependence with $\la$ (away from the QCP) and $L$ (close to QCP) for 1D Dirac model can be seen which is clearly not present in the 2D case in the susceptibility limit. The main figure verifies the logarithmic scaling with $L$ when $\la=0.01$. Inset shows the logarithmic dependence with $\la$ for $L=100$ in 1D case and no such dependence   for 2D case with $L_x=L_y=100$. In  the main figure as well as in the inset  $\del=0.001$ while  $k_{\rm max}=\pi$.}
\label{Fig:Wirr_scal2}
\end{figure}

Using the exact form  of $W_{\rm irr}^{T=0}$ as given in Eq.~(\ref{Wirr_anl}), one can now derive the scaling behavior of the
same in different limits: in the susceptibility limit 
$\de^{-1}$ is the largest length scale that never appears in the scaling.  In 
{this} limit, we then find:

\ba
 W_{\rm irr}^{T=0}/L &\sim& \del^2\log({2k_{max}}/{\lam});\del^{-\nu}>L>\lam^{-\nu}>k_{\rm max}^{-1}\nonumber\\ \label{Wirr_scala_lam}\\ 
&\sim& \del^2\log({Lk_{max}}/{2\pi}); \del^{-\nu}>\lam^{-\nu}>L>k_{\rm max}^{-1}\nonumber\\
\label{Wirr_scala_L}
\ea

On the other hand, in the thermodynamic limit where $L$ is the largest length scale, using Eq.~(\ref{Wirr_anl}), we get

\ba
 W_{\rm irr}^{T=0}/L &\sim& {\del^2 \log(2k_{\rm max}/\lam)};  L>\del^{-\nu}>\lam^{-\nu}>k_{\rm max}^{-1} \nonumber \\ \label{Wirr_scalb_lam}\\ 
&\sim& \del^2 \log(2k_{\rm max}/\del); L>\lam^{-\nu}>\del^{-\nu}>k_{\rm max}^{-1} \nonumber \\
\label{Wirr_scalb_del}
\ea

\begin{figure}
\includegraphics[width=7.5cm]{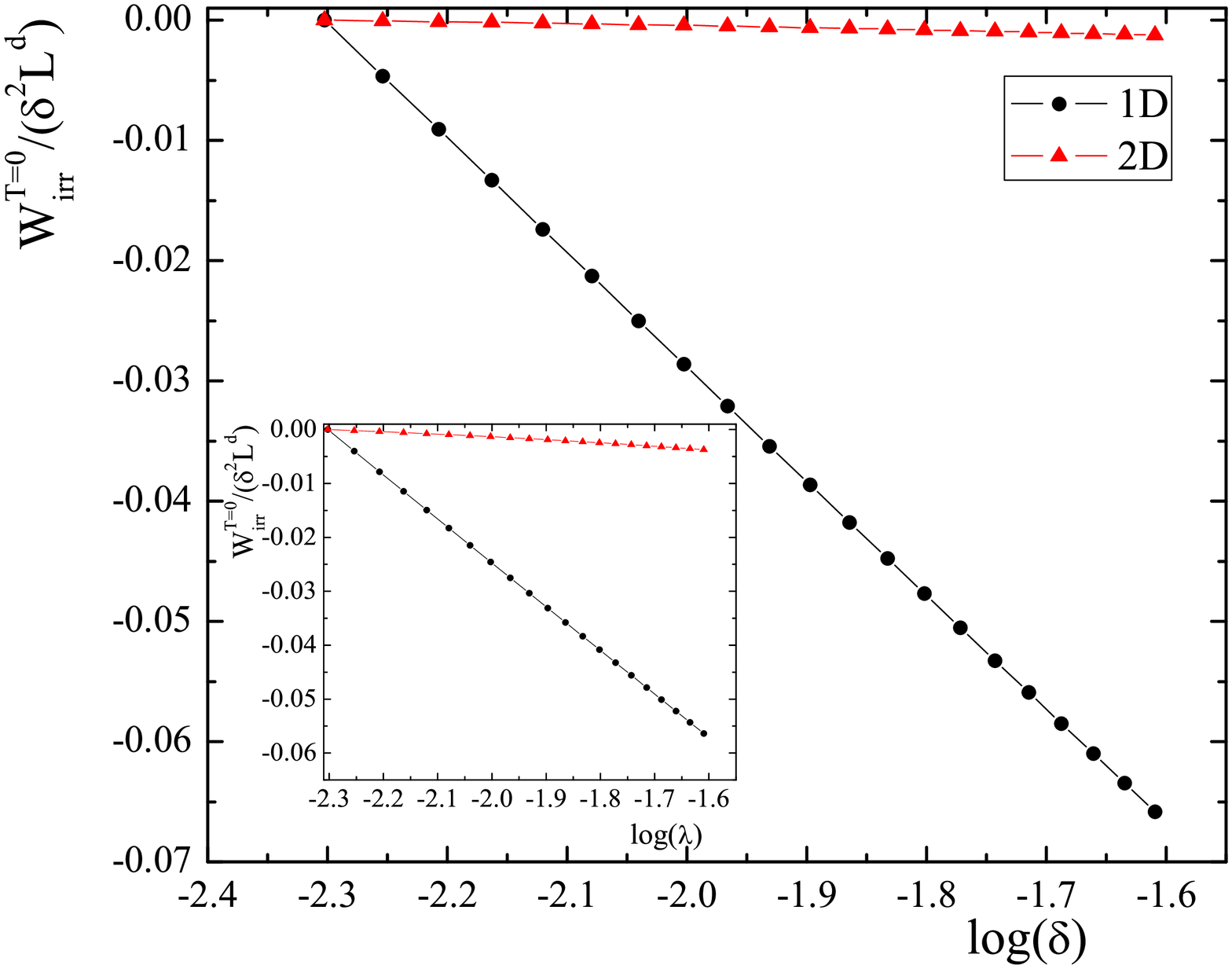}
\caption{ $W_{\rm irr}^{T=0}/(\del^2L^d)$ for 1D and 2D Dirac Hamiltonian verifies  the scaling proposed in the main text, showing a logarithmic scaling in the marginal situation (i.e. 1D Dirac model) which is absent in the 2D case. The main figure clearly confirms the scaling with $\de$ setting $\la=0.01$. Inset on the other hand shows the expected scaling away from the QCP in the thermodynamic limit with $\de=0.01$. The parameters for both the figures (main and inset) are $L=1000$ and $k_{\rm max}=\pi$.}
\label{Fig:Wirr_scal1}
\end{figure}

The scaling given in Eq.~(\ref {Wirr_scalb_lam}) might look identical to that given in Eq.~(\ref {Wirr_scala_lam}); however, we
would like to emphasize that these two relations are derived from the exact expression in the different limits. 
These scaling relations in the 1D and 2D cases
are illustrated  in Figs. (\ref{Fig:Wirr_scal2}) and (\ref{Fig:Wirr_scal1})  choosing the appropriate values of the parameters
in different limits.   We find that the leading scaling behavior of $W_{\rm irr}^{T=0}$ is given by $\de^2$; 
in the 1D case, there exist additional logarithmic corrections to the scaling with a scaling variable appropriate in the limit
under consideration. No such logarithmic correction is seen for the 2D case which can be established {using a similar
exact analytical form} of $W_{\rm irr}^{T=0}$.  Remarkably as we claimed earlier, in Fig.~(\ref{Fig:Wirr_scal2}),
the parameter $\del$ indeed {appears as}  a scaling variable in the thermodynamic limit when the system is close to the QCP; this is one of our most important findings which was never reported before.

Let us proceed to the finite temperature situation, and ask the question whether similar logarithmic
corrections are present in  $\Delta S_{\rm irr}$. The inherent $2 \times 2$ nature of the Dirac Hamiltonians enables us  to  calculate an exact expression for
$\Delta S_{\rm irr}$ also. For a single $k$ mode, one finds

\ba
\Delta {S_{\rm irr}}_{k}&=&\beta \left(E_k(\lam_0)+E_k(\lam_f)(1-2|\langle\psi_k^{-}(\lam_0)|\psi_k^{-}(\lam_f)\rangle|^2 \right)\non\\
&\times&\tanh\left({\frac{\beta E_k(\lam_0)}{2}}\right)
+2\log{\frac{\cosh(\beta E_k(\lam_f))}{\cosh(\beta E_k(\lam_0))}}.
\label{eq_sirr}
\ea
Integrating  the expression given in Eq.~(\ref{eq_sirr}) over all the momenta modes, one can derive the exact analytical forms of $\Delta S_{\rm irr}$ both
in 1D and 2D models. Analyzing those exact expressions, in the appropriate range of parameter values we can extract
the scaling behavior of $\Delta S_{\rm irr}$ which we have illustrated in Figs. (\ref{Fig:Sirr_scal1}) and (\ref{Fig:Sirr_scal2}).
What is remarkable is  that even though  the leading behavior of the scaling of $\Delta S_{\rm irr}$ is still given by $\de^2$ for both 1D
and 2D Dirac models, the logarithmic correction to the scaling persists in the 1D case for relatively low temperature while it
approaches the 2D case (when there is no logarithmic correction even at $T=0$) as thermal fluctuations increase. This
establishes our claim that the proposed scaling of $W_{\rm irr}^{T=0}$ indeed manifests in $\Delta S_{\rm irr}$ at low
temperature.

\begin{figure}[h!]
\centering
\subfigure[\  $\Delta S_{\rm irr}/L^d$ shows a logarithmic scaling with $\lam$ in 1D  Dirac with $L=100$; the slope approaches the 2D case
with increasing temperature where there is no logarithmic scaling. ]{
\includegraphics[width=4cm]{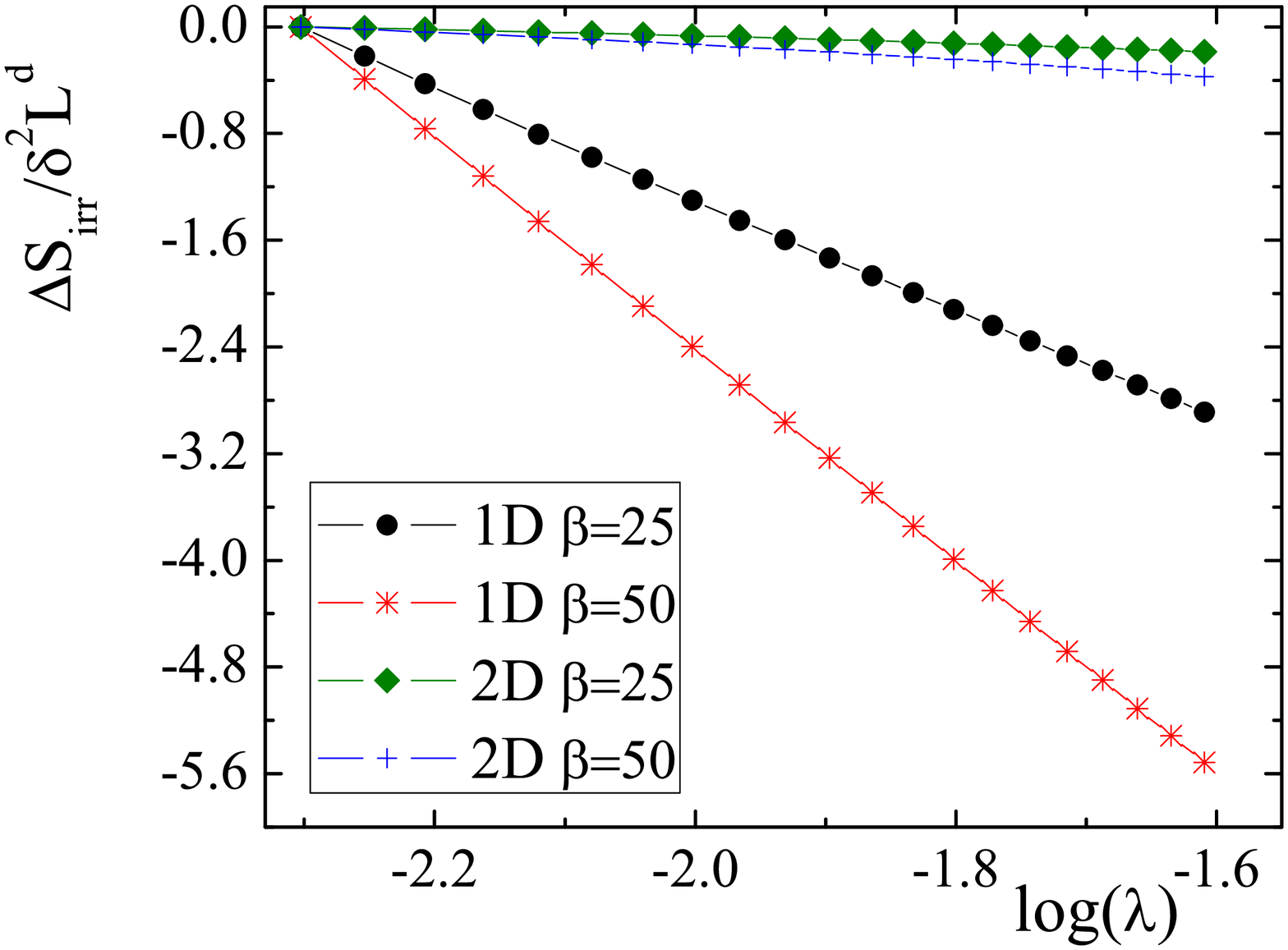}}
\subfigure[\ The same for the dependence with $L$ with $\lam=0.01$.  ]{
\includegraphics[width=4cm]{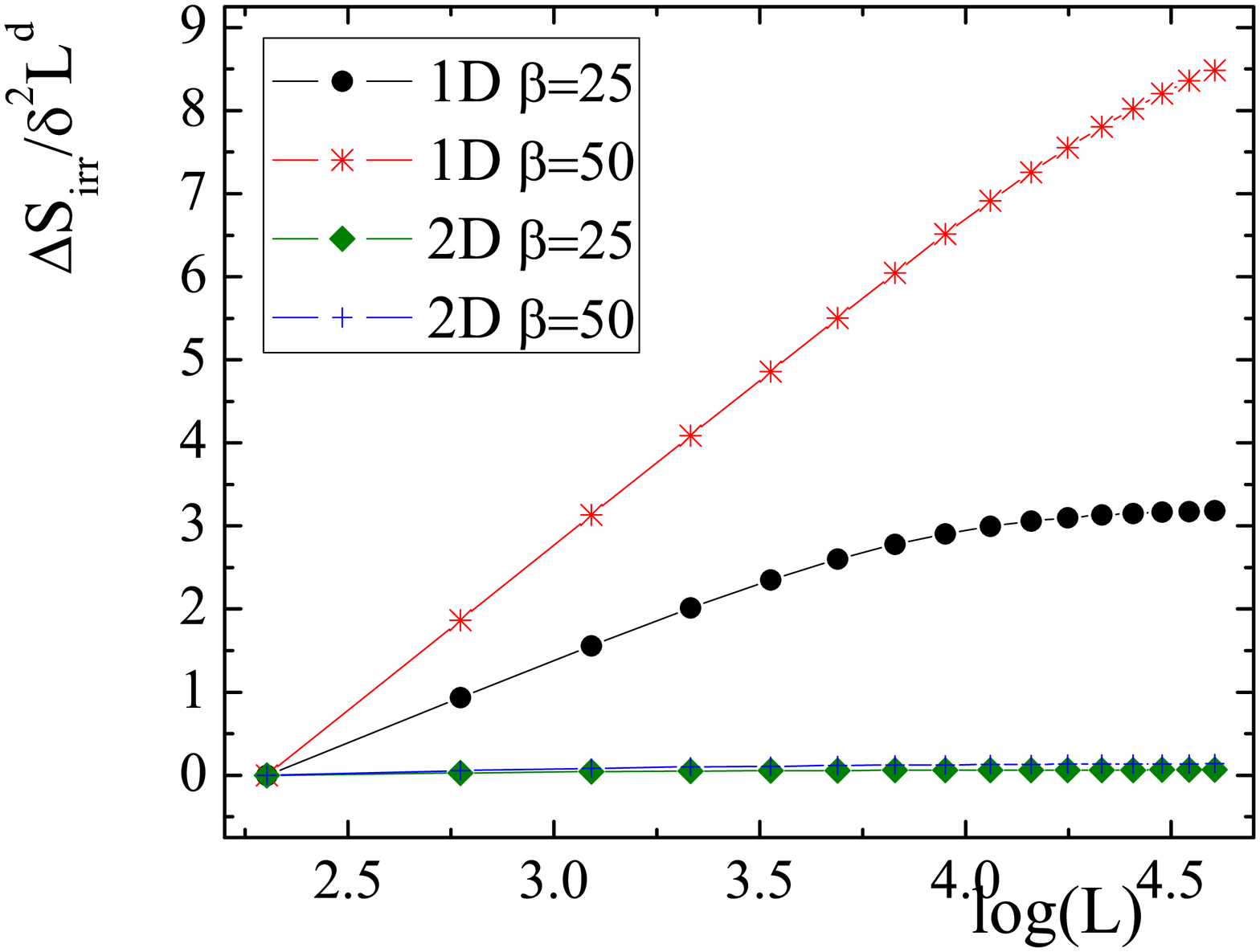}}
\caption{We show that in the susceptibility limit, the logarithmic scaling of  $\Delta S_{\rm irr}/L^d$ persists for low-temperature in the 1D Dirac case and its comparison with that of 2D Dirac where no such scaling is expected. We have taken $\de =0.001$ and $k_{\rm max} =\pi$.}
\label{Fig:Sirr_scal1}
\end{figure}

\begin{figure}[h!]
\centering
\subfigure[\ The scaling of $\Delta S_{\rm irr}/L^d$ in the thermodynamic limit, showing clear $\log$ dependence in 1D Dirac case which is clearly not observed in 2D Dirac case with $\de=0.01$.]{
\includegraphics[width=4cm]{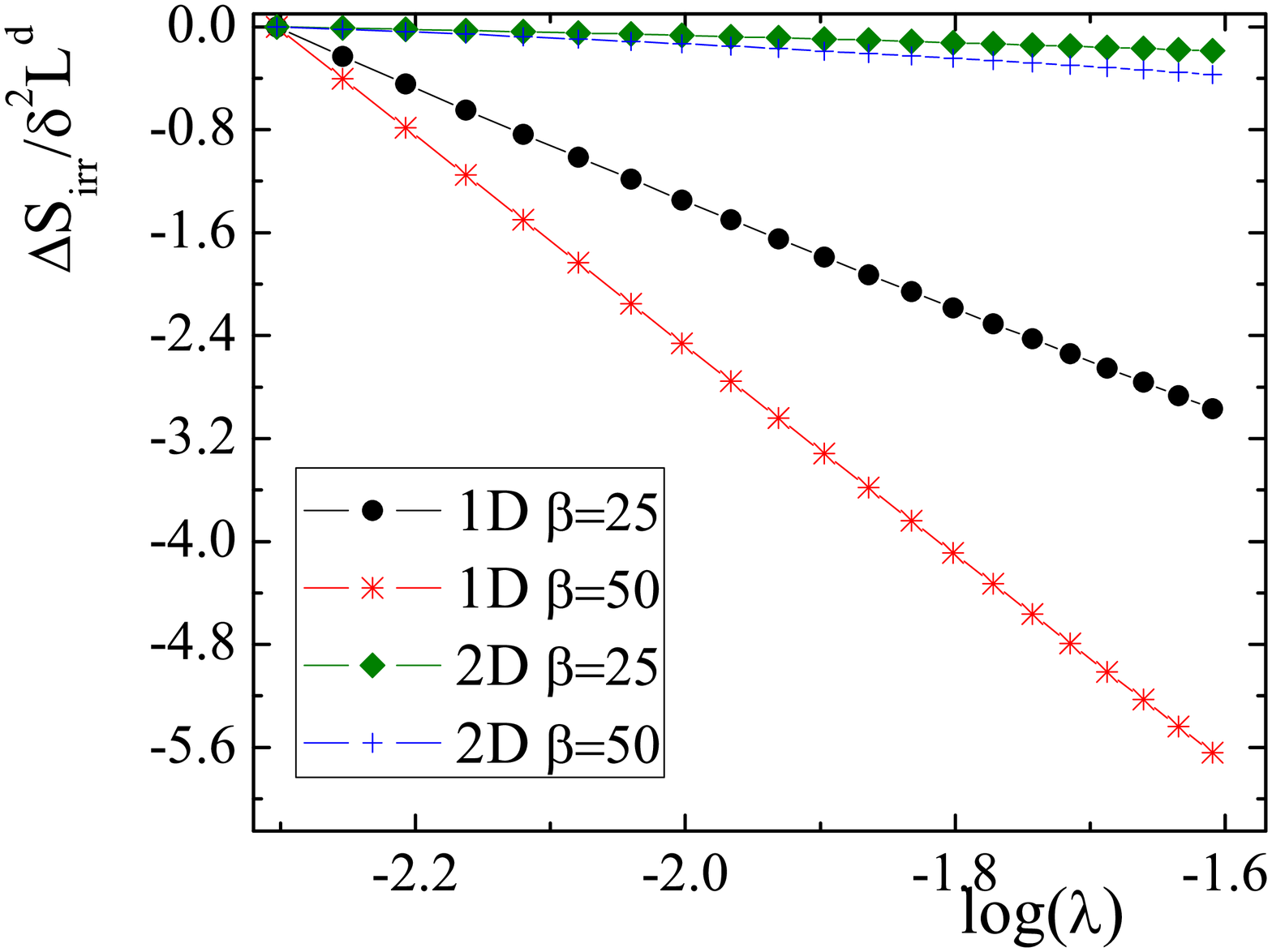}}
\subfigure[\ Scaling of $\Delta S_{\rm irr}/L^d$ with $\del$ shows a clear marginal behavior as a function of $\de$ (close to QCP) for 1D Dirac with $\la=0.01$.]{
\includegraphics[width=4cm]{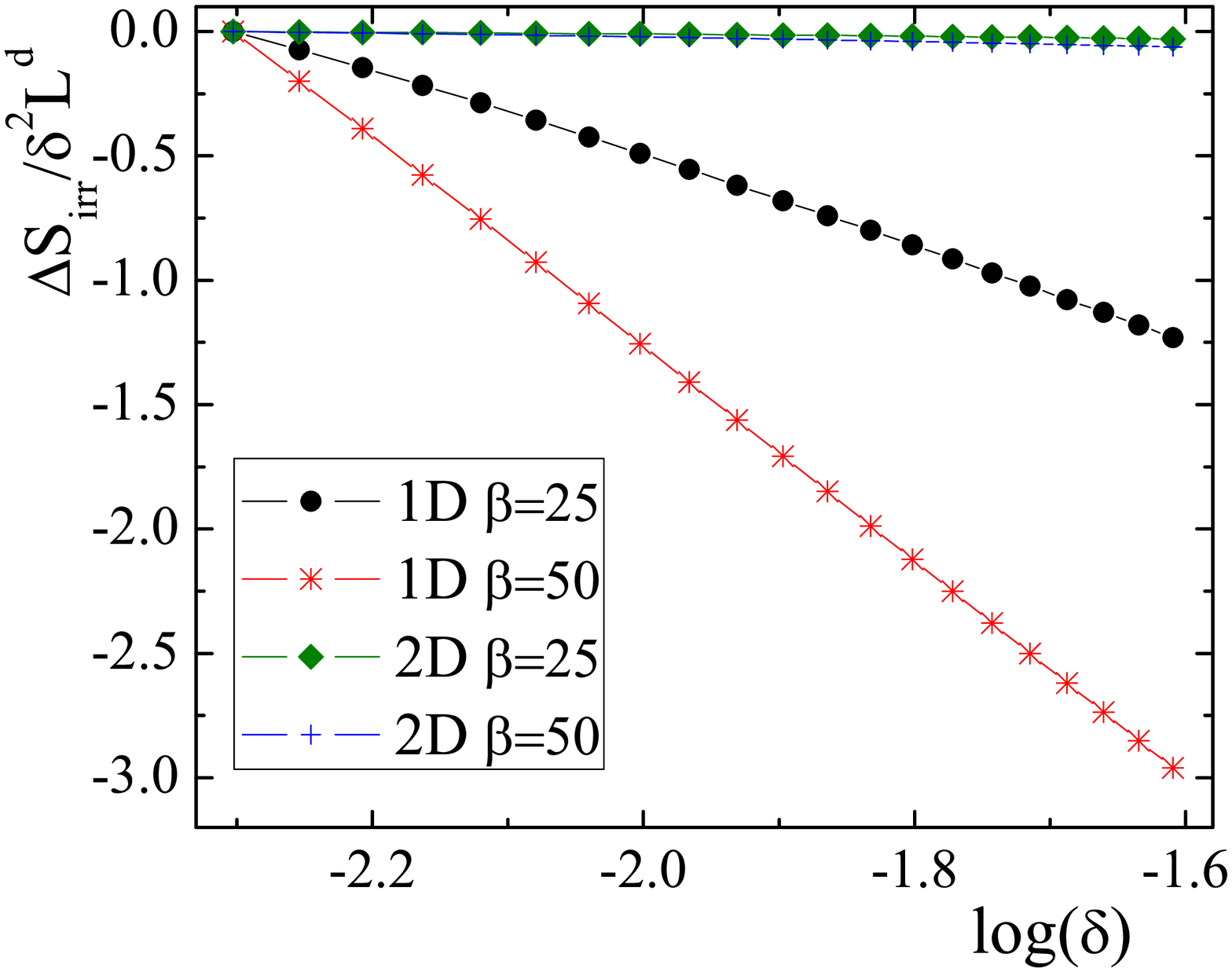}}
\caption{For 1D case there is a logarithmic correction to the scaling of $\Delta  S_{\rm irr}$ even in the thermodynamic limit which disappears at high temperatures
when the behavior is identical to that in the 2D case. Here, $L=1000$ and $k_{\rm max} = \pi$.}
\label{Fig:Sirr_scal2}
\end{figure}

Finally, we ask the question whether $\Delta S_{\rm irr}$ does always exhibit a sharp peak at the QCP when plotted
as a function of the tuning parameter $\la$ (see Fig.~(\ref{Fig:Sirr})); in 1D case indeed it does which  resembles the observation reported
in Fig.~(1) of the Ref. [\onlinecite{dorner12}]. However, we would like to draw attention to  the symmetric nature of $\Delta S_{\rm irr}$ on
either side of the QCP unlike the transverse Ising case.  On the contrary in the 2D case, there is no sharp peak even at sufficiently
low temperature. This leads us to the conclusion that above the marginal situation,  $\Delta S_{\rm irr}$ fails to be a good  indicator of
a QPT occurring at $T=0$.

%
\section{Concluding comments}

 \label{sec_conc}

In this paper, we have studied quantum and thermal quenches of a closed quantum system; in particular, our focus has
been restricted to the case when a parameter $\la$ of the Hamiltonian which is close to its QCP (at $\la=0$) is suddenly changed by an amount $\de$. Based on  the three length scales of the problem, namely $\la^{-\nu}$, $\de^{-\nu}$ and the system size $L$, we have
predicted the existence of two scaling regions, namely the susceptibility limit and the thermodynamic limit. Inspired by the
scaling of the ground state fidelity in these limits, we have proposed the scaling of $W_{\rm irr}^{T=0}$ in these limits both  close to the QCP and away from the QCP; furthermore,  a logarithmic correction with the parameter $\delta$ appears
in the scaling close to the QCP establishing that in the thermodynamic limit $\de$ obviously plays the role of a scaling variable.

We then asked the question whether these scaling relations survive in the case of a sudden quench in which
the initial state is a mixed state in thermal equilibrium with a heat bath and manifest in $\Delta S_{\rm irr}$.
 To address this question in a transparent way, we
have used 1D and 2D Dirac Hamiltonians where  arriving at the exact expressions for both $W_{\rm irr}^{T=0}$ and $\Delta S_{\rm irr}$ is indeed possible. These exact expressions are then analyzed to derive the scaling relations in different limits.
Remarkably, our study establishes the logarithmic correction (which is a signature of marginality)  appearing in the scaling of $W_{\rm irr}^{T=0}$ for 1D indeed survives
in the scaling of $\Delta S_{\rm irr}$ for low enough temperatures.  There is
no trace of any logarithmic corrections in the 2D case where the sharp peak in $\Delta S_{\rm irr}$ plotted as a function of $\lam$
also get broadened thereby diminishing its usefulness as an ideal detector of a QPT.

\begin{figure}[h!]
\centering
\subfigure[\ $\Delta S_{\rm irr}$ showing peak at the QCP ($\la=0$) for a 1D Dirac Hamiltonian. The peak height changes with the temperature keeping $\delta$ fixed to $0.01$]{
\includegraphics[width=4.0cm]{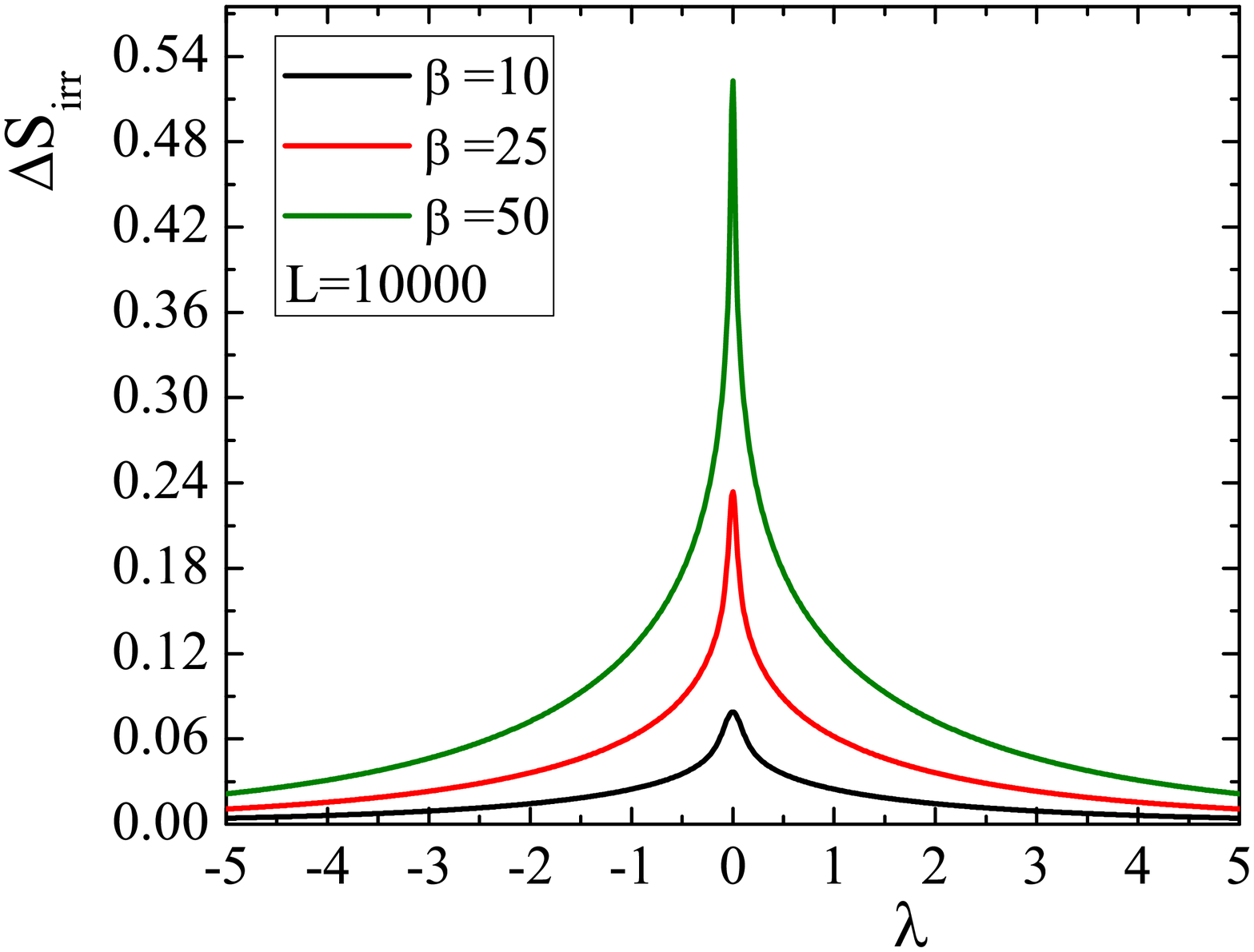}}
\subfigure[\ $\Delta S_{\rm irr}$ versus $\la$ shows peak at the QCP for 2D Dirac Hamiltonian for $\delta=0.01$ and changing $\beta$.]{
\includegraphics[width=4.0cm]{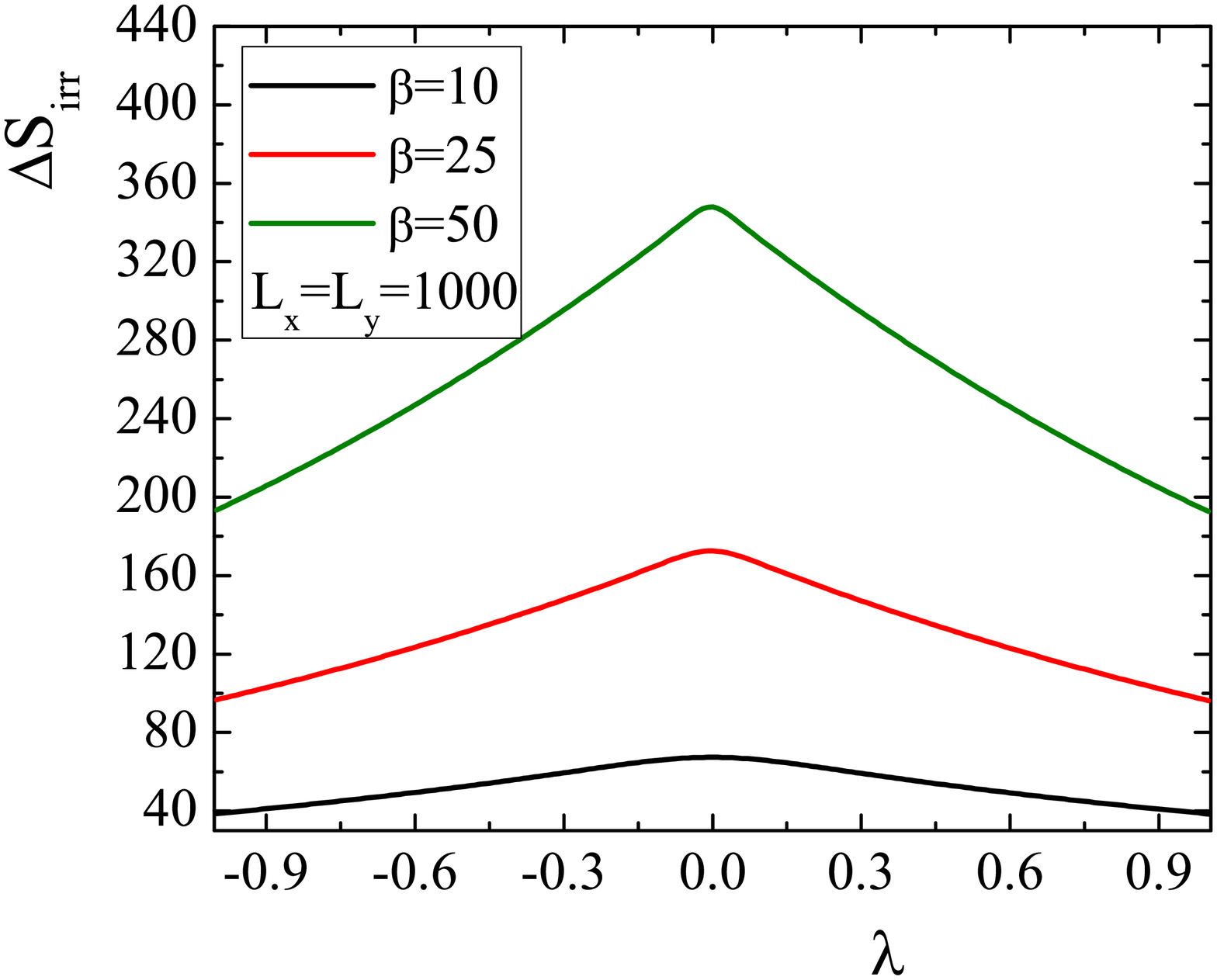}}
\caption{The figure above shows $\Delta S_{\rm irr}$ versus $\la$ for 1D and 2D Dirac Hamiltonians with different $\beta$ and $\delta=0.01$. There is a sharp peak at $\la=0$ for 1D which gets broadened in the 2D case. Here, $\beta$ increases from top
to bottom.}
\label{Fig:Sirr}
\end{figure}

\end{document}